\documentclass[epj]{svjour}

\usepackage[]{graphicx}
\usepackage{color}
\usepackage{float}

\renewcommand{\d}{{\rm d}}

\newcommand{\lamu}{{\lambda_{\rm u}}}
\newcommand {\E}  {{\varepsilon}}
\newcommand {\om} {{\omega}}
\newcommand {\Om} {{\Omega}}

\begin{document}

\title{Channeling and radiation of the $855$ MeV electrons enhanced by the re-channeling 
in a periodically bent diamond crystal}

\author{
Andrei V. Korol\inst{1}
\and 
Victor G. Bezchastnov\inst{2}
\and
Andrey V. Solov'yov\inst{1}\thanks{on leave from A.F. Ioffe Physical-Technical Institute, 
194021 St. Petersburg, Russian Federation}
}

\institute{
MBN Research Center, Altenh\"{o}ferallee 3, 60438 Frankfurt am Main, Germany
\and
A.F. Ioffe Physical-Technical Institute, 194021 St. Petersburg, Russian Federation
}

\mail{korol@mbnexplorer.com}

\authorrunning{A.V. Korol et al.}
\titlerunning{Channeling and radiation enhanced by the 
re-channeling in a periodically bent diamond crystal}

%\keywords{channeling; periodically bent crystal; de-channeling length; 
%radiation spectra}

\abstract{
Channeling properties and radiation spectra are studied on the grounds of 
numerical simulations for the $855$ MeV electrons in a periodically bent 
diamond crystal. The bent crystalline profiles are shown to enhance the 
re-channeling of the projectiles and to produce 
distinct lines in the radiation spectra. The results obtained are 
analyzed and contrasted to the properties of the planar channeling and 
of the channeling in uniformly bent crystals. 
}

\maketitle

\section{Introduction}

Propagation of relativistic charged particles in oriented crystals 
remain for several decades in focus of challenging research. 
As predicted by Lindhard~\cite{Lindhard1965}, the projectiles 
can channel in crystals moving along the crystalline planes or axes. 
The properties of channeling and the radiation produced by the projectiles 
have been receiving a significant interest, both with respect to a fundamental 
theory and the experiments, see, e.g., the 
book~\cite{ChannelingBook2014} and the references therein. 

In plane crystals, channeling enhances the radiation yield as compared 
to that produced by the non-channeling projectiles. 
The enhancement results from the inter-planar oscillations of the channeling 
projectiles and manifests itself in the radiation spectra at the energies 
determined by the oscillation frequencies and the projectile energies. 
The radiation from bent crystals displays additional features 
related to the bending. For example, the arc-bent crystals (also commonly 
referred to as uniformly bent), produce a synchrotron-type radiation 
at the energies dependent on the bending radius~\cite{BentSilicon_2014}. 
A periodically bent crystal can work as a {\em crystalline undulator} (CU) 
causing modulations 
in the motion of the projectiles with the frequencies determined by the bending 
period. This can result in {\em undulator lines} in the radiation spectra, 
which is an attractive property of CUs as the sources of monochromatic radiation 
in a sub-MeV to MeV energy range~\cite{KSG1998,KSG_review_1999}. A variety of 
experiments were performed to produce and detect the undulator radiation. The recent 
experimental studies as well as an outlook to the ongoing and future experiments can be 
found in, e.g, Refs.~\cite{BadEms_p38,BadEms_p58,BadEms_p63,Wienands_Talk_2016}. 

One can distinguish two basic types of periodical bending 
that lead to different relations between the frequencies of 
channeling and undulator oscillations of the projectiles. One is a 
large-amplitude long-period bending, originally 
suggested~\cite{KSG1998,KSG_review_1999} for the crystalline undulator 
as a device where the channeling particles follow the periodically bent 
crystalline planes. There, the undulator modulation frequencies are 
smaller than the frequencies of the channeling oscillations, and the 
undulator spectral lines arise at the energies below the energies of 
the channeling lines. Another type is a small-amplitude short-period 
bending~\cite{Kostyuk_PRL2013} which 
produces regular jitter-type modulations of the projectile motion with 
the period shorter than the period of the channeling oscillations. 
The corresponding lines arise in the radiation spectra at the energies 
exceeding the energies of the channeling peaks. 

Channeling and radiation in the bent crystals have been extensively studied, 
to be mentioned among others is a series of 
works~\cite{BentSilicon_2013,Sub_GeV_2013,Backe_EtAl_PRL2014,Multi_GeV_2014,Sushko_EtAl_NIMB_v355_p39_2015,Wistisen_EtAl_2016,Korol_EtAl_NIMB_v387_p41_2016}.
The most frequently addressed, both theoretically and experimentally, are the 
silicon crystals. In the present work, we study the channeling 
and radiation in diamond, motivated by the current experiments at the 
MAinzer MIcrotron (MAMI)~\cite{BadEms_p63,Backe_private}. In relation with the 
experimental conditions, we perform numerical simulations on the propagation of 
the $855$~MeV electrons through the plane, uniformly bent and periodically bent 
diamond (110). We focus on a large-amplitude long-period CU, in particular, 
compute the radiation spectra for the bending amplitude larger than the distance 
$d=1.26$~{\AA} between the (110) planes in the straight diamond. The simulations 
will show that the channeling electrons follow the bent planes and produce 
distinct undulator features in the radiation spectra. The results will also 
reveal an enhanced re-channeling process which develops in the periodically bent 
structures and makes them superior to the uniformly bent ones in supporting the 
channeling.

The theoretical framework of the simulations is described in detail in, e.g.,
Refs.~\cite{ChannelingBook2014,ChanModuleMBN_2013}.  
We use the \textsc{MBN Explorer} package~\cite{MBN_ExplorerPaper,MBN_ExplorerSite} 
to compute the motion of the ultra-relativistic projectiles through the crystals 
along with a dynamical simulation of the crystalline structures in the course of 
motion~\cite{ChanModuleMBN_2013}. The computations account for the interaction 
of the projectiles with the separate atoms of the environments. 
We remark that, with a variety of implemented inter-atomic potentials, 
the~\textsc{MBN Explorer} supports rigorous simulations of various environments, 
including crystalline, amorphous, and biological ones. 
The trajectories of the projectiles are used to compute the 
spectra of the emitted radiation. 
The simulation procedure is outlined in Section 2 and is followed by 
the studies of channeling properties (Section 3) and of radiation 
spectra (Section 4) for the diamond crystalline structures. 
The conclusions are given in Section 5.

\section{Simulation procedure} 

For numerical studies we introduce a reference frame with the $z$-axis along the 
incoming beam. The structure of the straight crystal 
is simulated with the (110)-planes parallel to the $(xz)$-plane. The crystal 
is oriented such that the beam is not directed along any major crystallographic axis, 
as this is done in the experimental setup in order to exclude the axial channeling. 
The $y$-axis corresponds to the inter-planar (transverse) 
direction. To simulate bent crystals, the coordinates $x',y',z'$ 
of each lattice node are obtained from the coordinates $x,y,z$ of the same node in 
the straight crystal as
\begin{equation} 
x' = x, \;\;\; 
y' = y + \delta(z), \;\;\;
z' = z, 
\label{bent_structure}
\end{equation}
where $\delta(z)$ is the shape of the bent (110)-planes. 
For the uniformly bent structures, the shapes are determined by the bending radius $R$, 
$\delta(z) = R - \sqrt{R^2-z^2}$. 
The periodical bending is simulated according to the cosine shape 
\begin{equation}
\delta(z) = a \cos (2\pi z / \lamu), 
\label{bending_periodical}
\end{equation}
where $a$ and $\lamu$ are bending amplitude and period, respectively. The 
(110)-planes for the straight and bent structures determine the channels for the 
channeling motion of the projectiles. For the electrons, the equations for the 
mid-lines of the channels in the $yz$-frame are $y = kd + \delta(z)$, where $k$ 
are the integer numbers and $d$ is the inter-planar distance. The boundaries 
separating the neighboring channels correspond to the relations 
$y = (k+0.5)d + \delta(z)$. 

Within the simulations, the incoming particles are statistically sampled 
with respect to the initial coordinates and velocities at the crystalline 
entrance. The initial transverse coordinates are selected 
randomly from a domain around the middle of a channel, with the domain size 
$\Delta x > 3d$, $\Delta y > 2d$. The initial velocity 
has the value determined by the energy of particles and is predominantly oriented 
in the $z$-direction, i.e. the initial transverse velocities that account for the 
beam emittance are small compared to the longitudinal velocity. The results of 
our simulations correspond to the zero emittance i.e. the zero initial 
transverse velocities. 

For each incoming electron, the propagation through the crystal is computed by 
a numerical integration of the classical relativistic equations of motion. 
The crystalline environment is simulated with account for the thermal 
fluctuations of the lattice with an amplitude corresponding to the temperature 
$300$~K. The statistical procedures in sampling the incoming projectiles and 
accounting for the thermal fluctuations allow one to regard the simulations as 
the Monte Carlo ones. They result in statistical ensembles of the trajectories 
that are used to analyze the channeling and to compute the radiation spectra. 

As in our previous studies, we assume the channeling to occur in the segments of 
a trajectory where the transverse motion is restricted by the neighboring channel 
boundaries whereas the sign of transverse velocity changes at least two times. 
The channeling segments alternate with the segments 
of non-channeling (over-barrier) motion, as a result of two ``complementary'' 
processes experienced by the projectiles - de-channeling and re-channeling 
(referred in some studies to as volume reflection and volume capture, respectively, 
see, e.g., Ref.~\cite{Wistisen_EtAl_2016}). 
To quantity the channeling properties, we introduce and calculate 
the acceptance ${\cal A}$, 
the penetration depth $L_{\rm p}$, and the channeling lengths $L_{\rm ch}$ and 
$L_{\rm tot}$. In addition, we compute the variations of the relative amounts 
of the channeling projectiles in the crystal along the beam direction. 

A trajectory is regarded as accepted when it starts with a channeling segment 
(referred to as a primary channeling segment) from the crystalline entrance, and the 
acceptance is defined as ${\cal A} = N_{\rm acc}/N_0$, 
where $N_{\rm acc}$ and $N_0$ are the numbers of the accepted and of the 
all simulated trajectories, respectively. The penetration depth $L_{\rm p}$ 
(also denoted as $L_{{\rm p}1}$ is some previous studies) is 
calculated as the mean longitudinal length of the primary channeling segments. 
To determine the channeling lengths, a sum longitudinal 
extension of the channeling segments is calculated for each trajectory 
which displays a channeling motion anywhere in the crystal. The extensions 
are averaged over the trajectories displaying channeling and over all the simulated 
trajectories, yielding the channeling length $L_{\rm ch}$ and $L_{\rm tot}$, 
respectively. These lengths have a meaning of the sum longitudinal distance 
passed in the channeling mode by a channeling projectile and by the entire beam. 
Notice that $L_{\rm ch}$ exceeds both $L_{\rm p}$ and $L_{\rm tot}$, and that the ratio 
$L_{\rm tot}/L_{\rm ch}$ is equal to the ratio of the amount of particles displaying 
channeling to the total number of particles in the simulated beam. 
The lengths $L_{\rm p}$, $L_{\rm ch}$ and $L_{\rm tot}$ are important characteristics 
of the channeling and radiation. For example, an excess of $L_{\rm ch}$ over 
$L_{\rm p}$ quantifies effectiveness of the re-channeling process in the course 
of the projectile's propagation. Larger $L_{\rm ch}$ values correspond to higher 
intensities of the channeling peaks in the radiation spectra. Larger values of 
$L_{\rm tot}$ imply the channeling peaks to be more distinct with respect to the 
broad-spectrum background radiation produced by the non-channeling projectiles. 

The variation of the amount of channeling particles along the crystal are 
represented by the two types of fractions, the primary channeling 
fraction $N_{{\rm ch}0}(z)/N_{\rm acc}$ and the channeling fraction 
$N_{\rm ch}(z)/N_{\rm acc}$. For each $z$, these fractions are determined 
by the numbers $N_{{\rm ch}0}(z)$ and $N_{\rm ch}(z)$ of the channeling 
projectiles in the primary channeling segment and anywhere in the crystal, 
respectively. The first fraction is solely contributed by these accepted 
projectiles that keep moving in the channeling mode until they experience the 
first de-channeling. The second fraction is additionally contributed by the 
projectiles (not necessarily accepted ones) that enter the channeling mode 
as a result of re-channeling.

The radiation produced by the projectiles is computed according to the 
quasi-classical formalism~\cite{Baier}. A spectral intensity of the radiation 
within the cone $\theta\leq \theta_0$ along the beam direction from is evaluated as 
\begin{eqnarray}
\frac{\d E(\theta\leq\theta_0)}{\hbar\d\om}
=
\frac{1}{N_0}
\sum_{n=1}^{N_0} 
\int\limits_{0}^{2\pi}
\d \phi
\int\limits_{0}^{\theta_0}
\theta \d\theta\,
\frac{\d^2 E_n}{\hbar\d\om\,\d\Om},  
\label{spectra}
\end{eqnarray} 
where the sum runs over the simulated trajectories of the total number $N_0$. 
The integration over the radiation angles is performed numerically. 
For the relativistic particles, the radiation is mostly concentrated within the cone 
$\theta \leq \gamma^{-1}$, where $\gamma$ is the Lorentz-factor corresponding to the 
beam energy. As in the previous simulations, we compute the radiation spectra 
for two different aperture values $\theta_0$, much smaller and much larger 
than $\gamma^{-1}$, respectively. 

The above described simulation procedure has been applied to study the channeling 
and radiation for the electrons with the energy $\E=855$~MeV passing through the 
straight, uniformly bent and periodically bent (110) diamond structures of the thickness 
$L=25~\mu$m in the beam direction. 
The numbers of the simulated trajectories 
varied between $5000$ and $9000$ and provided the results with statistical uncertainties 
$3.291\sigma$ with $\sigma$ being the standard deviations 
(i.e., the uncertainties indicated in the tables and plots correspond 
to a $99.9 \%$ confidence interval). 
For the periodical bending, we picked up a particular amplitude $a=2.5$~{\AA} and a 
period $\lamu=5~\mu$m of large-amplitude long-period diamond CU used in the experiments 
at MAMI. When referring to a CU in what follows, we will assume, if not explicitly 
specified otherwise, the diamond (110) sample with the above parameters. 

\section{Channeling properties}

We have first studied the effect of uniform bending on the acceptance and penetration 
properties of electrons. The values of ${\cal A}$, $L_{\rm p}$, $L_{\rm ch}$ and 
$L_{\rm tot}$ obtained for the straight and bent crystals are given in 
Table~\ref{Table_e-lengths.C}. For each value of the bending radius $R$, we have 
also calculated the bending parameter $C=\varepsilon/(RU_{\max}^{\prime})$ as the 
ratio of two opposite forces: the centrifugal force $\E/R$ and the stabilizing force 
$U_{\max}^{\prime}=6.7$~GeV/cm estimated as the maximal gradient of a continuous 
inter-planar potential in the diamond (110)~\cite{ChannelingBook2014}. 

\begin{table}[ht]
\caption{
Acceptance ${\cal A}$, penetration depths $L_{\rm p}$ and 
channeling lengths $L_{\rm ch}$ and $L_{\rm tot}$ for the 
$855$~MeV electrons passing through the $25~\mu$m thick
straight and uniformly bent diamond (110) crystal. The first two columns 
indicate the bending radius $R$ and parameter $C$. The first line 
corresponds to the straight crystal with $R=\infty$ and $C=0$.
}
\label{Table_e-lengths.C}
\resizebox{\columnwidth}{!}{ %%%
\begin{tabular}{rrrrrr}
\hline\noalign{\smallskip}
$R$ (cm)& $C$ & $\cal{A}$ & $L_{\rm p}\;\;(\mu$m) & $L_{\rm ch}\;\;(\mu$m) & $L_{\rm tot}\;\;(\mu$m) \\ 
\noalign{\smallskip}\hline\noalign{\smallskip}
$\infty$ & 0.00 & 0.73 & $12.01 \pm 0.40$ & $15.88 \pm 0.36$ & $14.72 \pm 0.33$ \\
6.11     & 0.02 & 0.73 & $11.70 \pm 0.46$ & $14.53 \pm 0.48$ & $12.01 \pm 0.41$ \\ 
2.44     & 0.05 & 0.67 & $11.02 \pm 0.46$ & $11.83 \pm 0.56$ & $ 8.42 \pm 0.41$ \\ 
1.74     & 0.07 & 0.62 & $10.58 \pm 0.47$ & $11.13 \pm 0.64$ & $ 7.10 \pm 0.42$ \\ 
1.22     & 0.10 & 0.57 & $10.06 \pm 0.47$ & $10.31 \pm 0.66$ & $ 6.03 \pm 0.39$ \\  
0.81     & 0.15 & 0.50 & $ 9.12 \pm 0.48$ & $ 9.23 \pm 0.75$ & $ 4.67 \pm 0.38$ \\ 
0.61     & 0.20 & 0.45 & $ 7.81 \pm 0.40$ & $ 7.89 \pm 0.66$ & $ 3.57 \pm 0.29$ \\ 
0.49     & 0.25 & 0.40 & $ 7.06 \pm 0.38$ & $ 7.13 \pm 0.69$ & $ 2.87 \pm 0.26$ \\ 
0.41     & 0.30 & 0.36 & $ 6.39 \pm 0.32$ & $ 6.40 \pm 0.61$ & $ 2.33 \pm 0.21$ \\ 
0.39     & 0.35 & 0.32 & $ 5.84 \pm 0.37$ & $ 5.86 \pm 0.79$ & $ 1.86 \pm 0.22$ \\ 
0.31     & 0.40 & 0.30 & $ 5.25 \pm 0.29$ & $ 5.26 \pm 0.66$ & $ 1.57 \pm 0.17$ \\ 
0.24     & 0.50 & 0.24 & $ 4.32 \pm 0.29$ & $ 4.33 \pm 0.79$ & $ 1.02 \pm 0.15$ \\
\noalign{\smallskip}\hline
\end{tabular}
} %%%
\end{table}

\begin{figure}[ht]
\centering
\includegraphics[width=1.00\columnwidth]{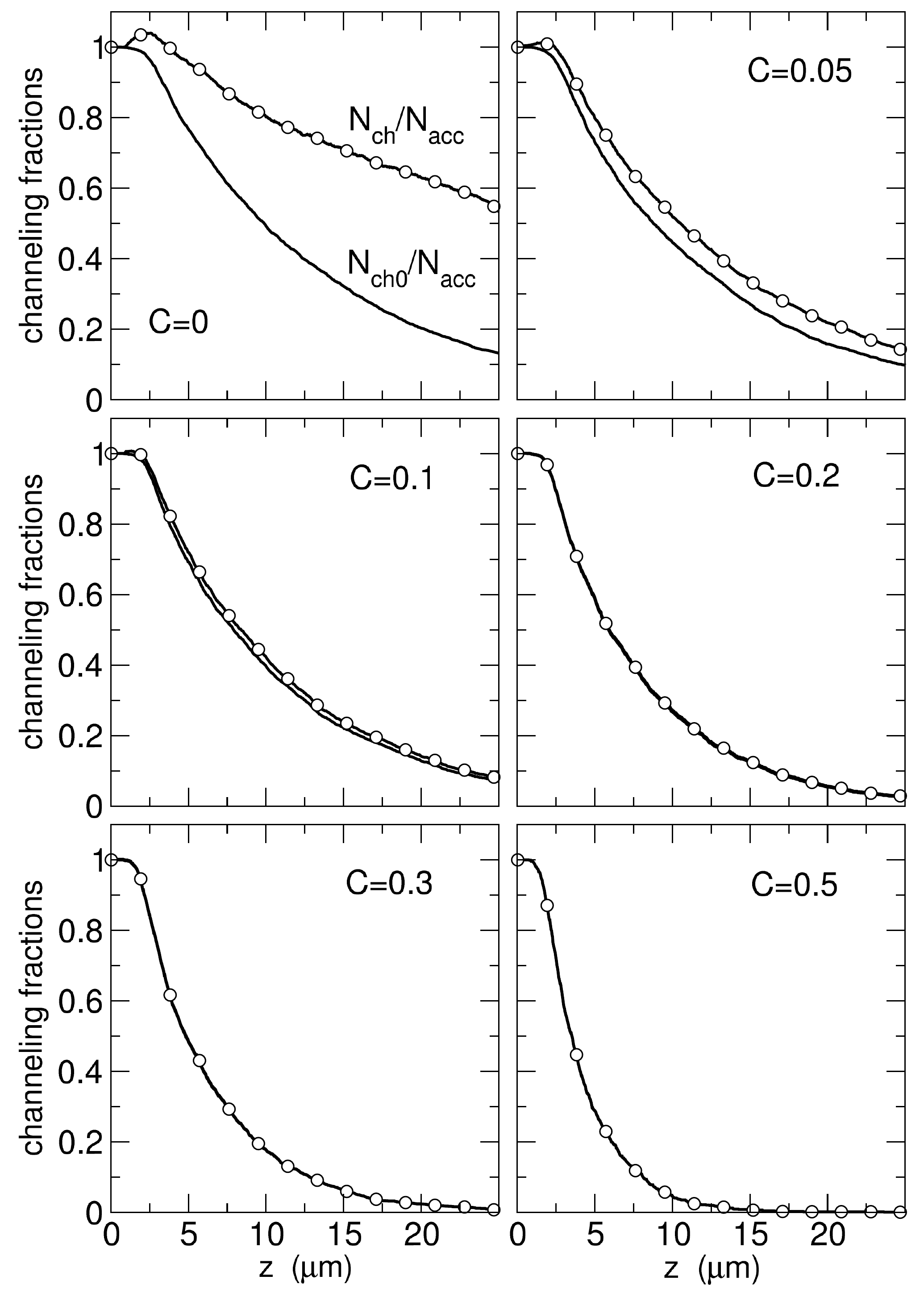}
\caption{Fractions of the channeling electrons in the uniformly bent diamond (110) 
for different values of the bending parameter $C$ ($C=0$ corresponds to the straight crystal). 
The solid lines show the primary fractions $N_{\rm ch0}/N_{\rm acc}$, while 
the dots connected by the lines show the fractions $N_{\rm ch}/N_{\rm acc}$ accounting for 
the re-channeling. The difference between the two types of fractions becomes imperceptible 
for the values $C>0.1$. 
} 
\label{e-bent-channeling}
\end{figure}

The acceptance and lengths are maximal for the straight crystal ($R=\infty$, $C=0$). 
With increasing bending radius, they gradually decrease. 
For the values of bending parameter not exceeding $0.1$, the channeling lengths exhibit 
statistically significant excess over the penetration depths, 
indicating thereby an effectiveness of 
the re-channeling process. For the bending with larger $C$, the re-channeling events become 
rare as reflected by close values of $L_{\rm p}$ and $L_{\rm ch}$. The total channeling 
length $L_{\rm tot}$ decreases with stronger rate as compared to decrease of 
$L_{\rm p}$ and $L_{\rm ch}$. The latter behavior reflects deteriorating possibilities 
for the projectiles to enter the channeling regime of motion. We notice that there are 
no peculiarities in decreasing the acceptance and the lengths and can conclude 
that increasing bending suppresses channeling along overall crystal, i.e. just as in 
the primary channeling segments so also in the following segments that may result 
from the re-channeling. Such an effect of the bending on the channeling properties is 
additionally demonstrated by the plots of the channeling fractions 
(see Figure~\ref{e-bent-channeling}). At any $z$ along the beam path in the crystal, 
both the amounts of the primary channeling electrons (solid lines in the plots) and the 
amounts including the re-channeled projectiles (dots connected by the lines) are 
smaller for stronger bending (larger values of $C$). An excess of the number $N_{\rm ch}$ 
of the electrons that channel both upon entering the crystal as well as having experienced 
the re-channeling over the number $N_{\rm ch0}$ of the primary channeling electrons is 
most prominent for the straight crystal (see the plot for $C=0$). 
With increasing $C$, this excess diminishes and for $C>0.1$ becomes hardly observable 
indicating the re-channeling to not develop in any noticeable extent.   

Next we turn to the channeling properties for the periodically bent 
diamond. Table~\ref{Table_e-5micron-lengths-CU} presents the values 
${\cal A}$, $L_{\rm p}$, $L_{\rm ch}$ and $L_{\rm tot}$ 
obtained from the simulations for the bending period $\lamu=5~\mu$m and 
amplitudes $a$ varying from zero (the reference 
case of a straight crystal) to $5.5$~{\AA}. 
To compare the results with the these for the uniform bending, 
it is instructive to refer them to the values for the bending parameter. For the 
periodical profile~(\ref{bending_periodical}) the curvature 
\begin{equation}
\frac{1}{R} \simeq \left| \frac{{\rm d}^2\delta(z)}{{\rm d}z^2} \right| 
            = \frac{4\pi^2a}{\lamu^2} \left| \cos \left( \frac{2\pi z}{\lamu} \right) \right|
\label{curvature_periodical}
\end{equation}
as a function of $z$ varies with the period $\lamu/2$ between the zero and 
maximum $(1/R)_{\rm max} = 4\pi^2a/\lambda_{\rm u}^2$ values, with 
$\langle 1/R \rangle = 8\pi a/\lambda_{\rm u}^2$ being the mean curvature. 
Notice that Eq.~(\ref{curvature_periodical}) does not account for the slop 
${\rm d}\delta/{\rm d}z \propto a/\lamu$ which is small compared to unity. 
For more flexibility in relating the periodical bending to the uniform one, 
we have included in the table both 
the mean $\langle C \rangle = (\E/U^\prime_{\rm max})\langle 1/R \rangle$ and 
the maximal $C_{\rm max} = (\E/U^\prime_{\rm max})(1/R)_{\rm max}$ values for 
the bending parameter. 

\begin{table}[ht]
\caption{ 
Acceptance ${\cal A}$ and characteristic lengths $L_{\rm p}$, $L_{\rm ch}$ and 
$L_{\rm tot}$ for the propagation of the $855$~MeV electrons 
through a $25~\mu$m thick diamond (110) undulator. The crystal is periodically 
bent with the period $\lamu=5~\mu$m and different amplitudes $a$ indicated in the 
table, $\langle C \rangle$ and $C_{\rm max}$ are the mean and maximal values for 
the related bending parameter (see the text). 
The first line corresponds to the straight crystal. 
}
\label{Table_e-5micron-lengths-CU}
\resizebox{\columnwidth}{!}{ %%%
\begin{tabular}{rrrrrrr}
\hline\noalign{\smallskip}
$a$ (\AA) & $\langle C \rangle$ & $C_{\rm max}$ & ${\cal A}$ & $L_{\rm p}\;\;(\mu$m) & 
$L_{\rm ch}\;\;(\mu$m) & $L_{\rm tot}\;\;(\mu$m) \\
\noalign{\smallskip}\hline\noalign{\smallskip}
0.0 & 0.00 & 0.00 & 0.73 & $ 12.01\pm0.40$ & $ 15.88\pm0.36$ & $ 14.72\pm0.33$ \\
0.5 & 0.06 & 0.10 & 0.62 & $ 10.10\pm0.39$ & $ 13.80\pm0.32$ & $ 13.00\pm0.30$ \\ 
1.0 & 0.12 & 0.19 & 0.53 & $  8.34\pm0.39$ & $ 11.33\pm0.29$ & $ 10.81\pm0.27$ \\ 
1.5 & 0.19 & 0.29 & 0.44 & $  6.96\pm0.33$ & $  9.29\pm0.34$ & $  8.73\pm0.23$ \\ 
2.0 & 0.24 & 0.38 & 0.38 & $  5.90\pm0.27$ & $  7.72\pm0.20$ & $  6.94\pm0.18$ \\ 
2.5 & 0.31 & 0.48 & 0.33 & $  5.18\pm0.23$ & $  6.60\pm0.17$ & $  4.55\pm0.15$ \\
3.0 & 0.37 & 0.58 & 0.28 & $  4.39\pm0.24$ & $  5.71\pm0.19$ & $  4.66\pm0.16$ \\ 
3.5 & 0.43 & 0.68 & 0.22 & $  3.84\pm0.16$ & $  4.93\pm0.13$ & $  2.94\pm0.10$ \\ 
4.0 & 0.49 & 0.77 & 0.19 & $  3.52\pm0.15$ & $  4.41\pm0.12$ & $  2.64\pm0.09$ \\ 
4.5 & 0.55 & 0.87 & 0.15 & $  3.18\pm0.14$ & $  4.00\pm0.12$ & $  2.47\pm0.08$ \\ 
5.0 & 0.62 & 0.97 & 0.12 & $  2.86\pm0.12$ & $  3.55\pm0.12$ & $  1.95\pm0.07$ \\ 
5.5 & 0.68 & 1.06 & 0.08 & $  2.60\pm0.15$ & $  3.27\pm0.20$ & $  1.49\pm0.09$ \\ 
\noalign{\smallskip}\hline
\end{tabular}
} %%%
\end{table}

The acceptance and the lengths in Table~\ref{Table_e-5micron-lengths-CU}, being maximal 
for the straight crystal, decrease with increasing the bending amplitude $a$ and 
related values $\langle C \rangle$ and $C_{\rm max}$ of the bending parameter. 
Such a behavior is similar to that studied for the bending with constant radius, 
however for the periodical bending the channeling length $L_{\rm ch}$ remains 
larger than the penetration depth $L_{\rm p}$, even for the largest 
amplitude $a=5.5~\mu$m and corresponding values $\langle C \rangle = 0.676$ and 
$C_{\rm max} = 1.061$. Notice that the two latter estimates of the bending 
parameter substantially exceed the value $C=0.1$ beyond which the re-channeling 
appears to be completely suppressed in the uniformly bent crystals. Therefore, 
in the periodically bent structures, the increasing curvature 
(reflected by the increasing mean and maximal values for the parameter $C$) 
appears to not deteriorate the channeling as strong as in the 
arc-shape structures. In fact, the periodical bending can even enhance 
the re-channeling process, as we discuss below. 

\begin{figure}[ht]
\centering
\begin{tabular}{l}
\hspace{-0.4cm}
\includegraphics[width=1.00\columnwidth]{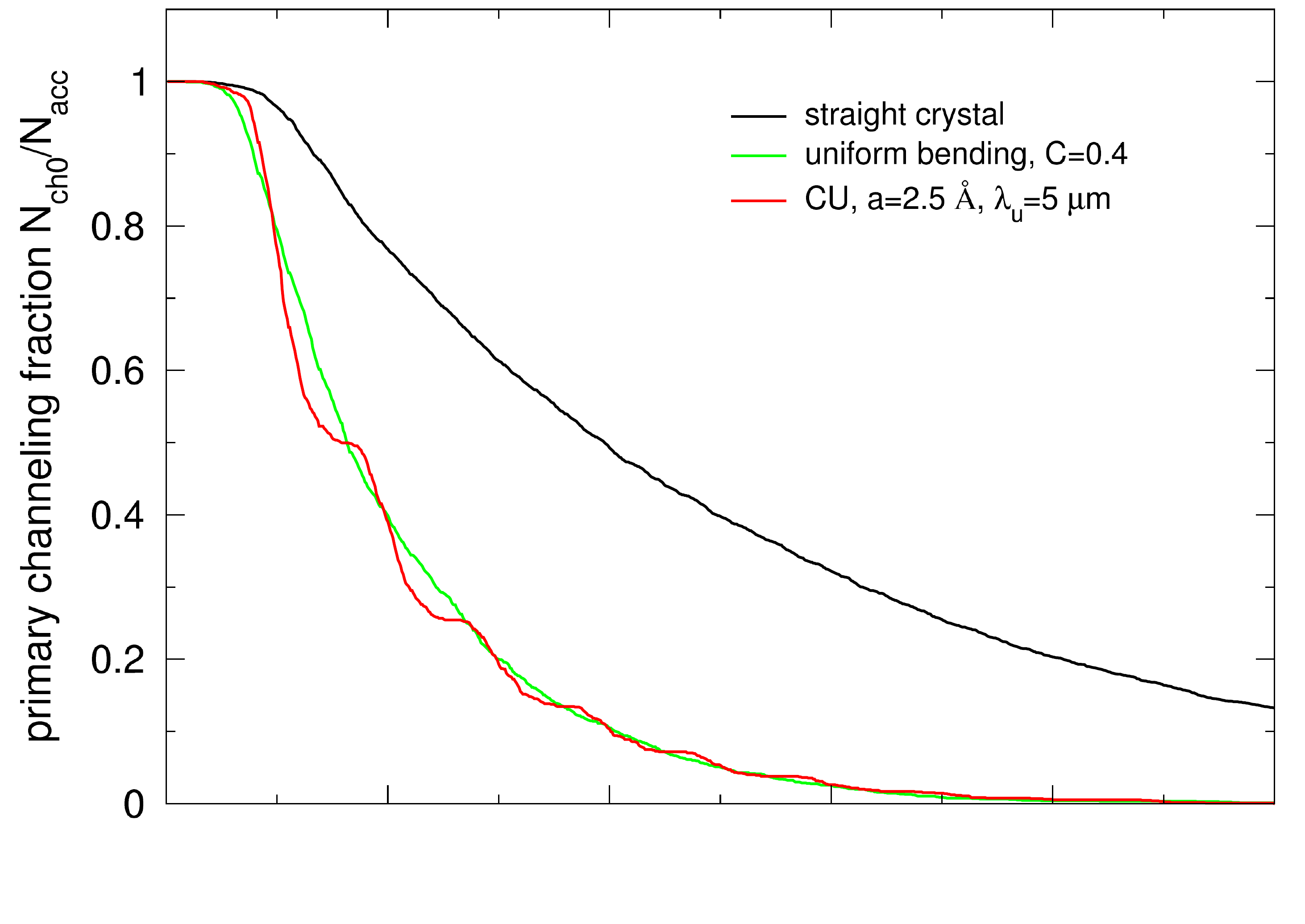} 
\vspace{-0.82cm}
\\
\hspace{-0.4cm}
\includegraphics[width=1.00\columnwidth]{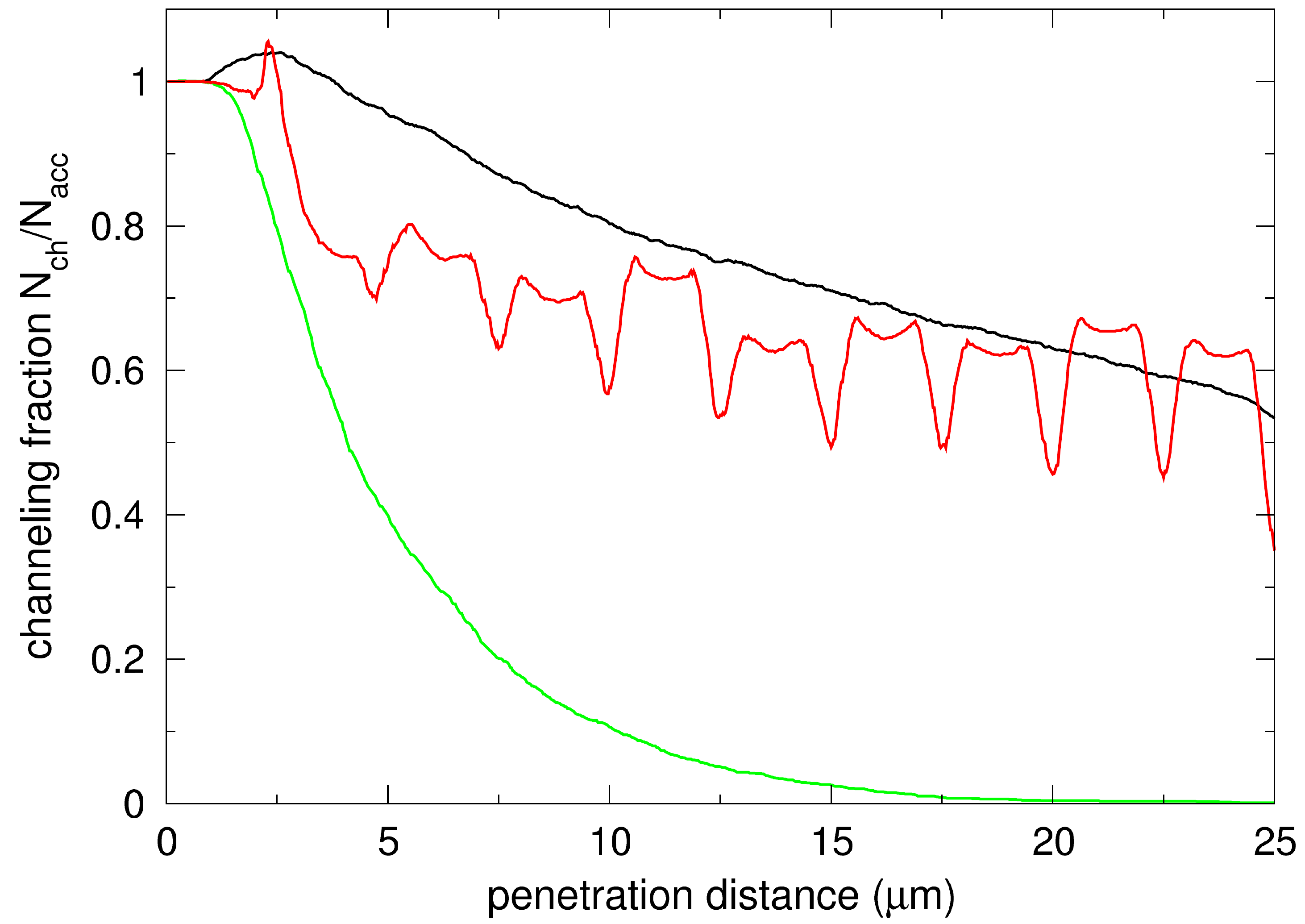}
\end{tabular}
\caption{Fractions of the channeling electrons in the 
straight, uniformly bent ($C=0.4$) and periodically bent ($a=2.5$~{\AA}, $\lambda_u=5~\mu$m) 
diamond (110). Top: primary fractions. Bottom: fractions with account for the re-channeling. 
} 
\label{e-channeling-compare}
\end{figure}

For a particular bending amplitude $a=2.5~\mu$m we have calculated the fractions of 
the electrons displaying a channeling motion at the distance $z$ from the crystalline 
entrance. They are presented in Fig.~\ref{e-channeling-compare} and compared with 
the fractions in the straight and arc-shape crystals. We have found the primary 
fraction in the crystal bent with a constant radius and $C=0.4$ 
to reflect well a smoothed behavior of the primary fraction in the periodically 
bent crystal, see the right plot in the figure. 
The selected constant bending parameter gets into the gap between 
$\langle C \rangle$ and $C_{\rm max}$ for the periodical bending. We can therefore 
conclude that the periodical bending influences the channeling in the primary 
channeling segment in a same way as for an arc-shape crystal with a $C$-value 
in between the mean and maximum values for this parameter for the 
periodical structure. In other words, the de-channeling from the primary channeling 
segment develops in the periodically bent crystal in a similar way as in the 
arc-shape crystal. 

As already reflected in Table~\ref{Table_e-lengths.C} by the close values of $L_{\rm p}$ 
and $L_{\rm ch}$, the re-channeling in the arc-shape crystal does not develop effectively, 
and the two channeling fractions (shown by the green curves in Fig.~\ref{e-channeling-compare}) 
do not exhibit any noticeable difference from each other. In contrast, we encounter a drastically 
different character for the re-channeling in the undulator. 
The fraction of the electrons that move in the channeling mode along the periodically 
bent (110) planes (either upon entering the crystal or having experienced the re-channeling) 
significantly exceed the fraction of electrons channeling right upon entering the crystal, 
see the red curves in the figure. Furthermore, the fraction accounting for the re-channeling 
undergoes remarkable oscillations with varying the penetration distance (see the right plot). 
It is appealing that after passing a distance of $20~\mu$m, the electrons display the channeling 
fraction with the oscillation maxima even exceeding the fraction of channeling electrons in the 
straight crystal.  

The enhancement of the re-channeling and significant oscillations in the number of channeling 
electrons in the undulator are the major findings of our studies on the channeling properties. 
These effects clearly result from the periodical bending. The oscillations of the channeling 
fraction in the right plot of Fig.~\ref{e-channeling-compare} exhibit the samples of the double 
maxima followed by the minima. Along the entire length $25~\mu$m of the udulator, we encounter 
ten such oscillation samples that clearly follow one another with the period 
$\lamu/2 = 2.5~\mu$m of the varying curvature, Eq.~(\ref{curvature_periodical}). 
The positions of minima and double maxima correspond to the distances $z$ that are even and 
odd multiplies of $\lamu/2$, respectively. We can conclude that, at the distances corresponding 
to the maximal curvature, the effect of de-channeling is maximal leading to the minima of the 
channeling fraction. In contrast, at the distances where the curvature approaches the zero value, 
the re-channeling becomes most effective and yields a significant increase in the number of 
channeling electrons. 

It is also remarkable that an average decrease of the channeling fraction with increasing path 
along the undulator, being rather sharp at $z < 10~\mu$m, becomes significantly slower for 
larger $z$. This can be interpreted as a cumulative effect of re-channeling, since when 
passing a longer distance the electrons experience more times the zero curvature and encounter 
situations ``optimal'' to trigger the non-channeling motion to the channeling one. 
We remark that according to the diffusion theory of 
channeling, at sufficiently large distances the decrease of the channeling fraction becomes 
a power-low one, as it is well displayed in the right plot of the figure by the fractions 
for the straight crystal and by the average behavior of the fraction in the undulator.   

\begin{figure}[ht]
\centering
\includegraphics[width=1.00\columnwidth]{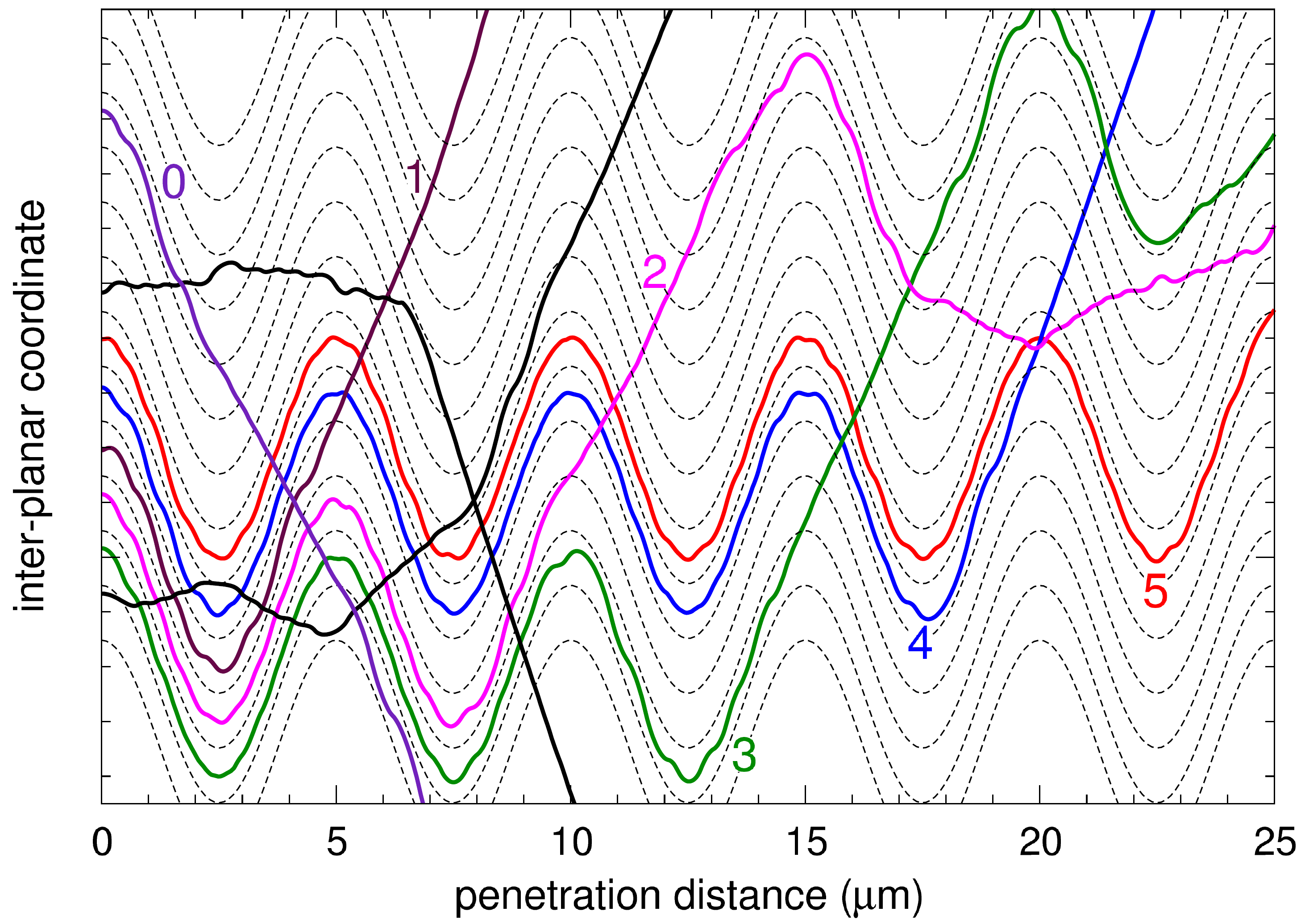}
\caption{
Representative electron trajectories in CU. 
Dashed lines visualize the electron channels. The accepted trajectories are labeled 
by the numbers $0,1,2,\dots,5$ of the periods which the electrons pass in the course of 
channeling motion from the CU entrance. Shown with no number labels are two trajectories 
for the not-accepted electrons.
}
\label{e-5micron-trajectories.fig}
\end{figure}

To complete the analysis of channeling in the periodically bent diamond, 
we present in Fig.~\ref{e-5micron-trajectories.fig} a set of simulated 
electron trajectories. We have included in the figure the trajectories with 
the channeling motions starting from the crystalline entrance and following a few 
bending periods, up to the maximal five periods (i.e. passing in the channeling mode 
through the entire crystal). These trajectories demonstrate that the channeling motion 
indeed develops in the large-amplitude long period diamond undulator. 
The figure also shows the trajectories for not-accepted projectiles as well 
as a trajectory for an accepted particle that stays in the channeling mode along 
a path shorter than the undulator period. The re-channeling is displayed in the figure 
by two trajectories that exhibit first the channeling motion (one trajectory - along 
two and another trajectory - along three undulator periods from the crystalline entrance). 
Notice that for the latter 
trajectories the channeling segments following the re-channeling are shorter than 
$\lamu$ (in fact, the lengths of these particular segments slightly exceed $\lamu/2$). 
We will refer to the types of trajectories presented in Fig.~\ref{e-5micron-trajectories.fig} 
when discussing the radiation spectra produced in the undulator. 

\section{Radiation spectra}

\begin{figure}[ht]
\centering
\begin{tabular}{l}
\hspace{-0.4cm}
\includegraphics[width=1.00\columnwidth]{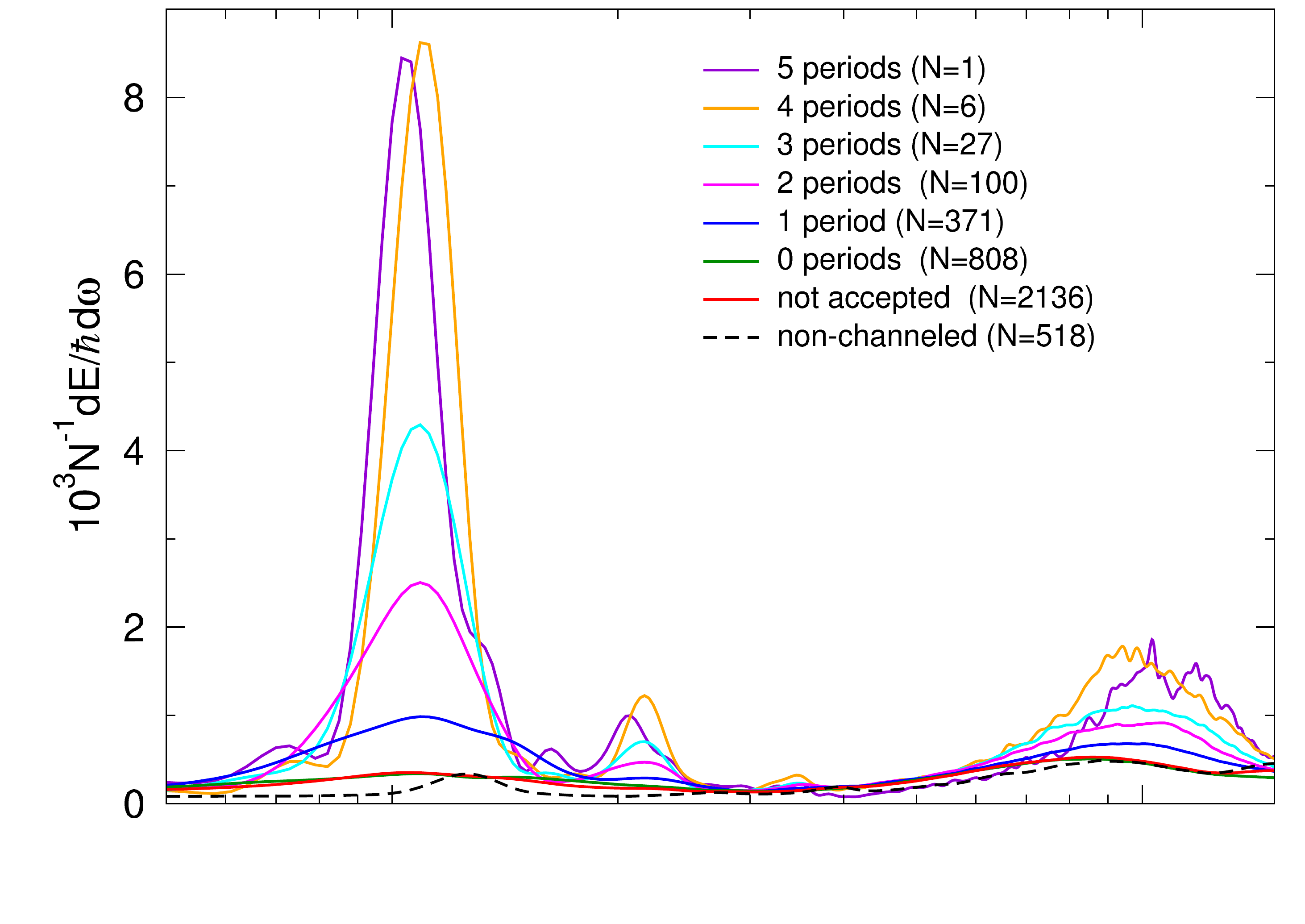} 
\vspace{-0.82cm}
\\
\hspace{-0.4cm}
\includegraphics[width=1.00\columnwidth]{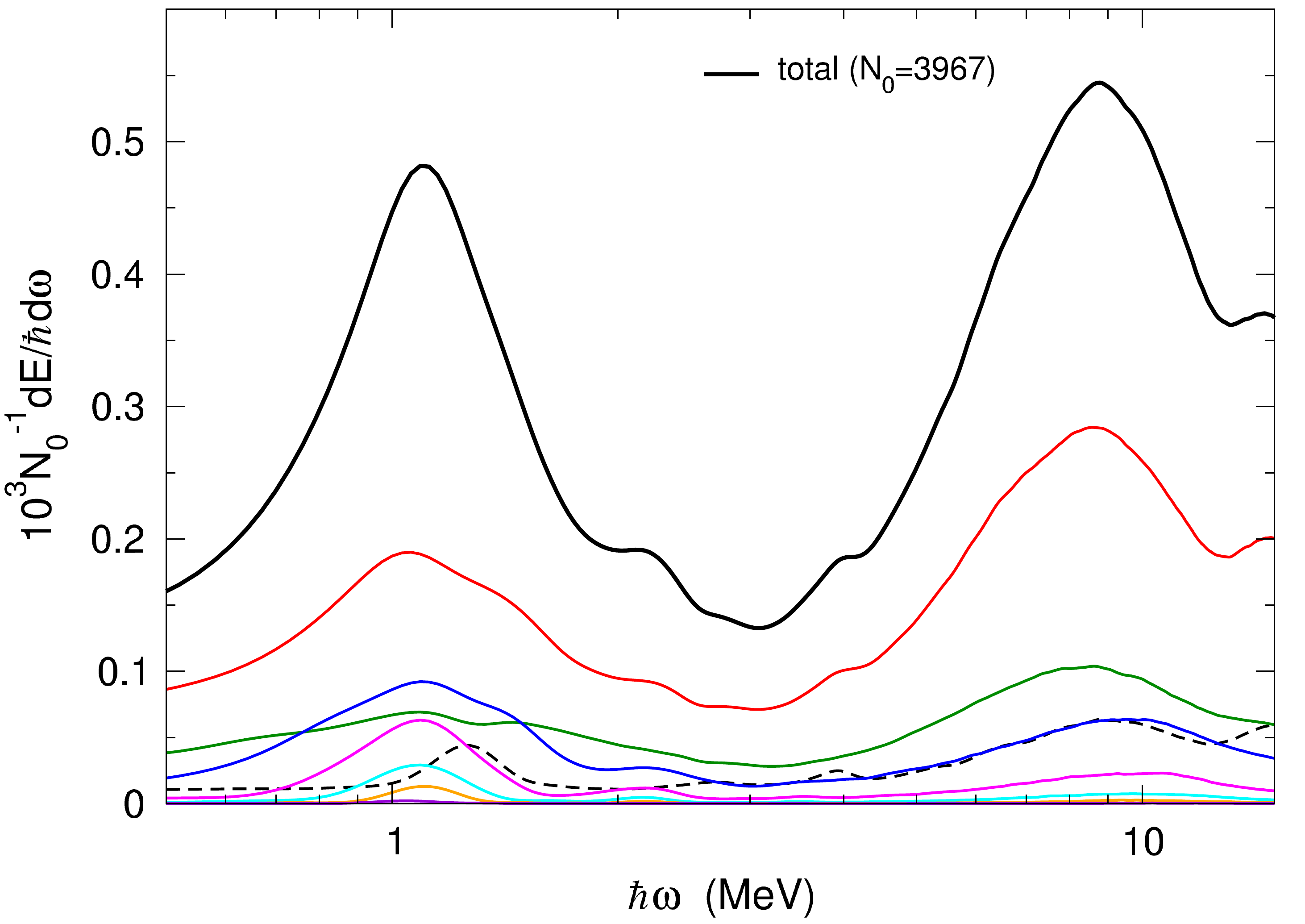}
\end{tabular}
\caption{Radiation spectra for $\theta_0=0.24$~mrad produced by the $855$~MeV electrons 
passing through CU. Top: spectra resulting from the trajectories of 
different types as indicated. The curves are normalized with respected to the numbers $N$ 
of the trajectories they are contributed by, i.e. they represent 
the spectra ``per single trajectory'' of a given type. Bottom: spectra resulting from the 
specific-type trajectories are shown normalized with 
respect to the total number $N_0$ of all the simulated trajectories. The 
corresponding curves represent the absolute contributions of the specific-type 
trajectories to the total emission spectrum shown by the uppermost bold curve.
} 
\label{e-spectra-partial}
\end{figure}

\begin{figure}[hb]
\centering
\begin{tabular}{l}
\hspace{-0.4cm}
\includegraphics[width=1.00\columnwidth]{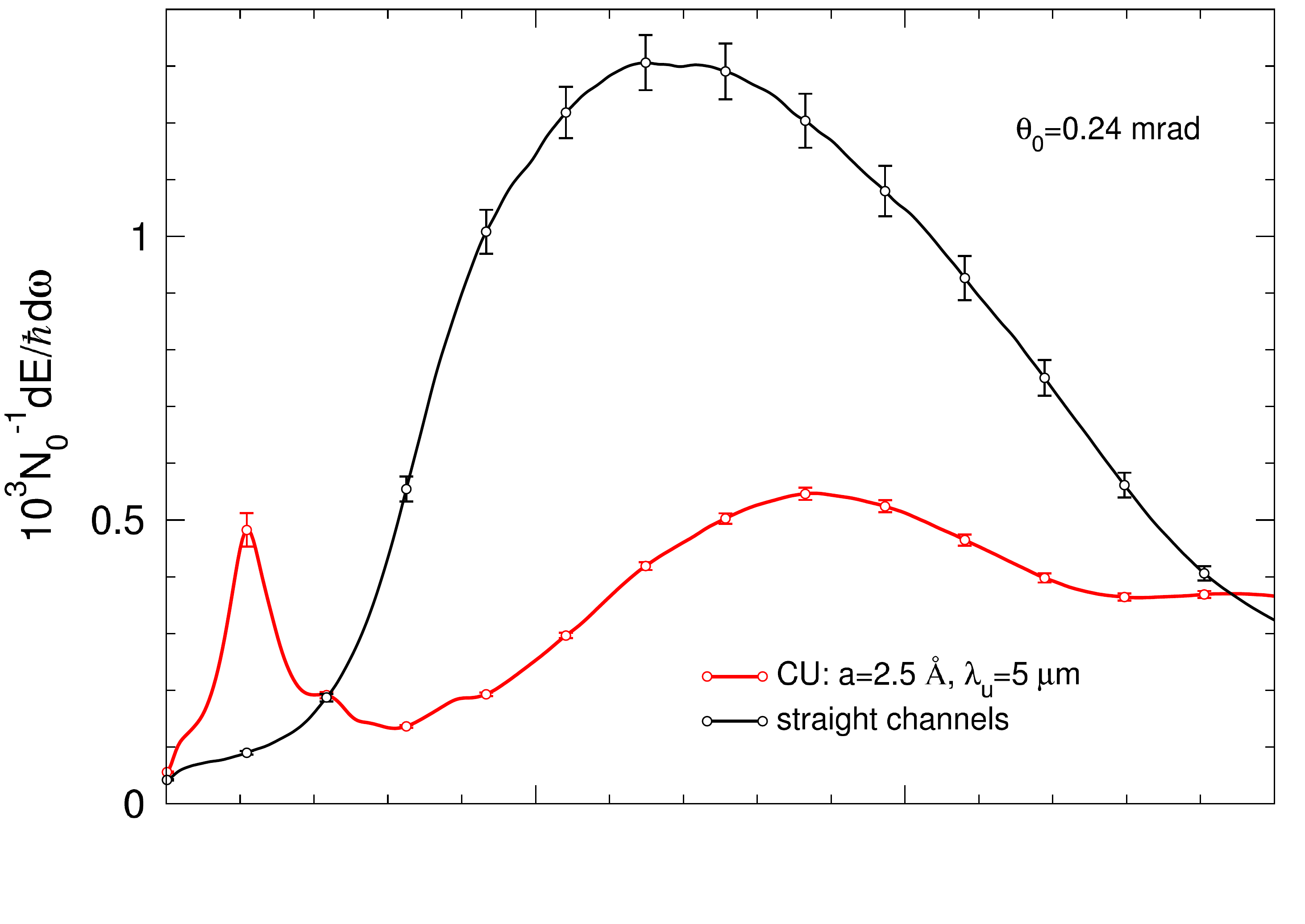} 
\vspace{-0.82cm}
\\
\hspace{-0.4cm}
\includegraphics[width=1.00\columnwidth]{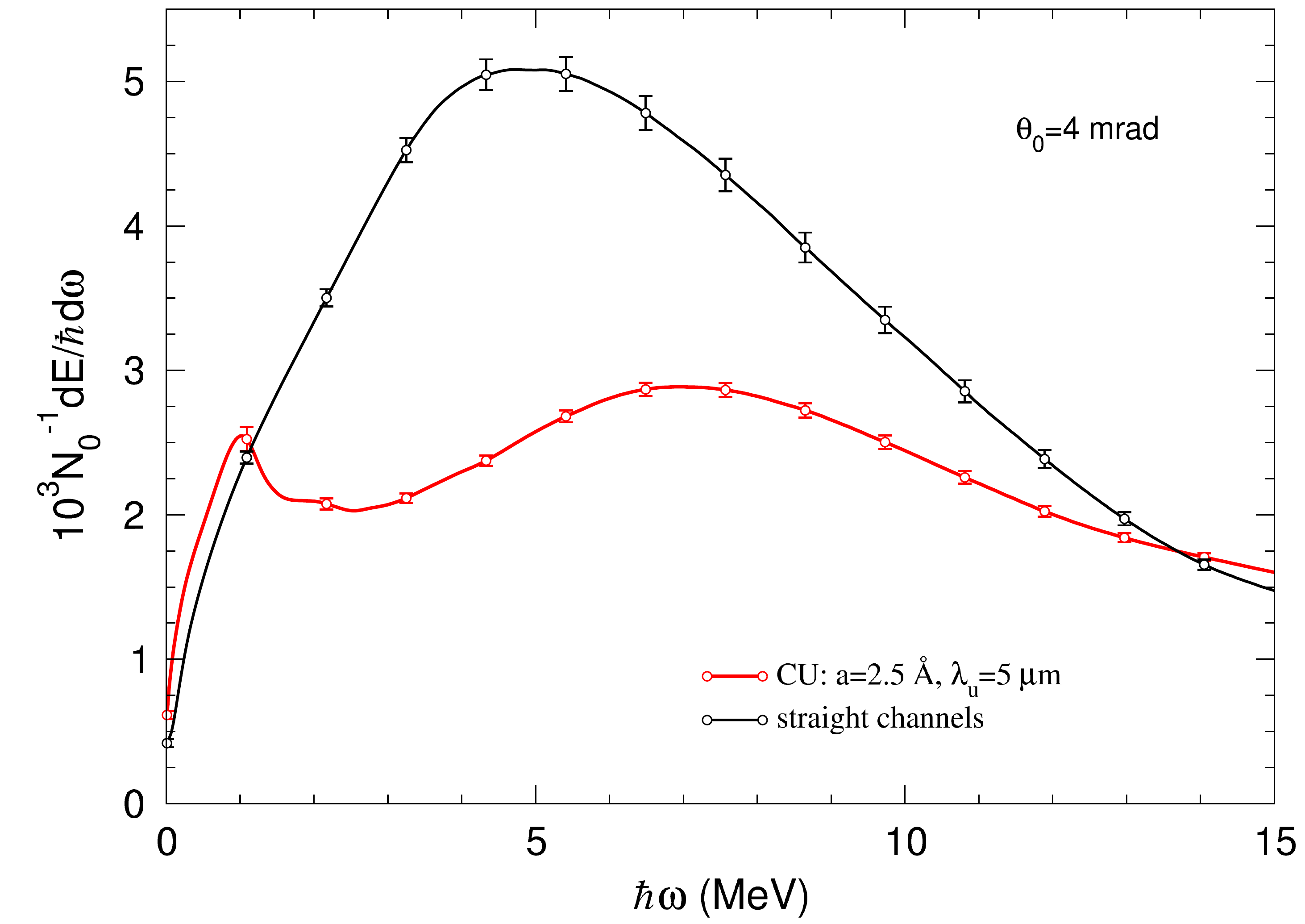}
\end{tabular}
\caption{Emission spectra produced by the $855$~MeV electrons in the CU (red curves) and a straight 
crystal (black curves). Top and bottom plots show the spectra for the emission aperture 
$\theta_0=0.24$~mrad and $\theta_0=4$~mrad, respectively. 
} 
\label{e-5micron-vs-straight_dE.fig}
\end{figure}

\begin{figure}[ht]
\centering
\begin{tabular}{l}
\hspace{-0.4cm}
\includegraphics[width=1.00\columnwidth]{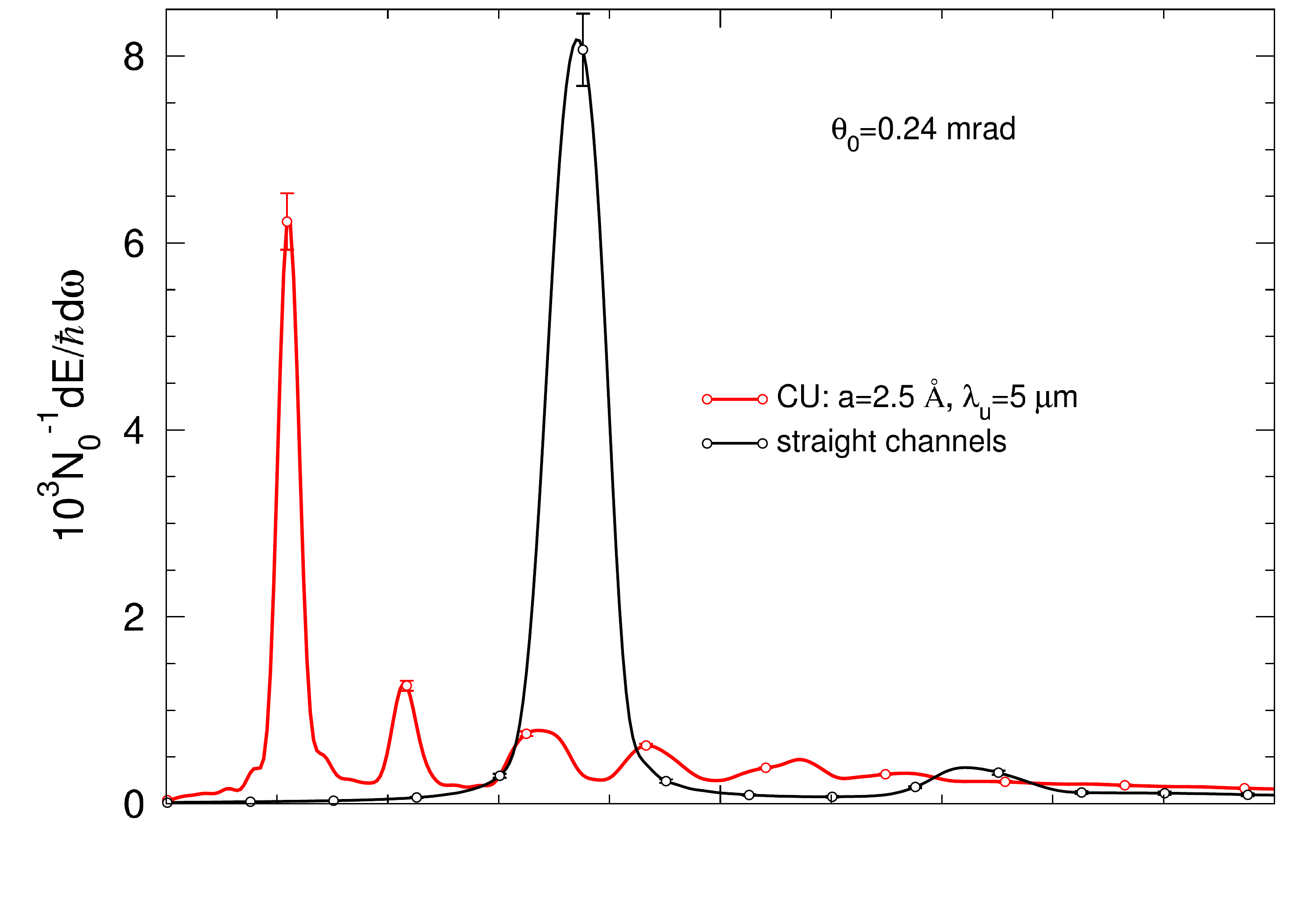} 
\vspace{-0.82cm}
\\
\hspace{-0.4cm}
\includegraphics[width=1.00\columnwidth]{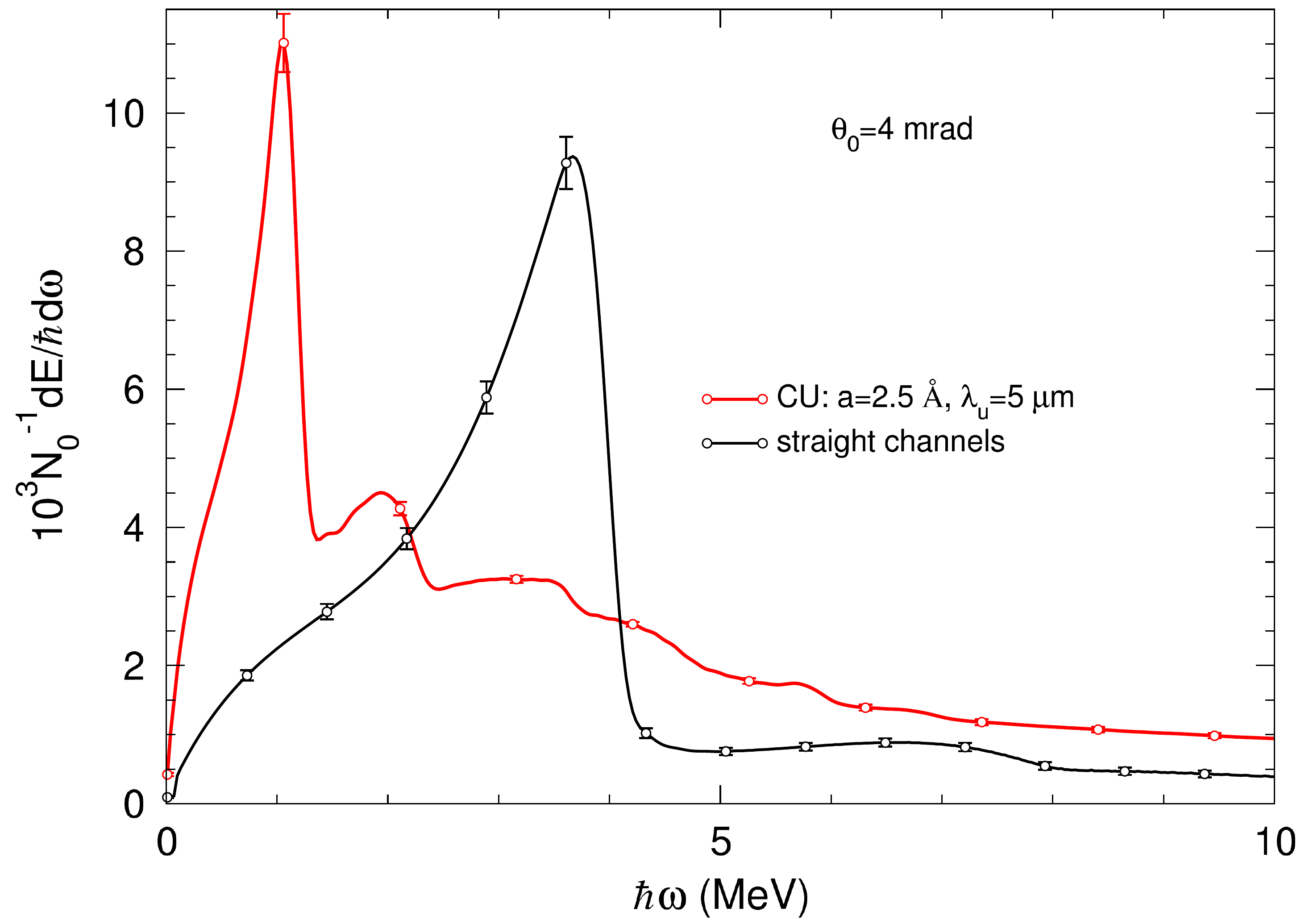}
\end{tabular}
\caption{Emission spectra produced by the $855$~MeV positrons for the same conditions as 
studied in Fig. \ref{e-5micron-vs-straight_dE.fig} for the electrons.
} 
\label{p-5micron-vs-straight_dE.fig}
\end{figure}

For the MAMI electron beam energy $855$~MeV, the natural opening angle for the radiation 
produced by the projectiles is $\gamma^{-1} \approx 0.6$~mrad. The aperture 
half-width for a detector used in the experiments is 
$0.24$~mrad~\cite{Backe_EtAl_PRL2014,Backe_private,Backe_EtAl_2011,Backe_EtAl_PRL_115_025504_2015}. 
We have applied Eq.~(\ref{spectra}) to calculate the emission spectra for 
$\theta_0=0.24$~mrad which is close to $\gamma^{-1}$, and for $\theta_0=4$~mrad 
which significantly exceeds the natural opening angle. The results are presented in 
Figs.~\ref{e-spectra-partial} and \ref{e-5micron-vs-straight_dE.fig}. 
Fig.~\ref{p-5micron-vs-straight_dE.fig} shows the spectra that can be produced by the positrons 
under the same conditions as for the electrons. 

To elucidate the formation of the undulator line in CU, we have computed spectra produced 
by the electrons moving along specific trajectories. 
We have selected, from the total amount of $3967$ simulated trajectories, different groups of 
trajectories of a same kind.  
It is natural to expect that the more periods an electron moves in the channeling mode upon 
entering the crystal, the more intensive udulator line it produces in the radiation spectra. 
Therefore, we have calculated the spectra for trajectories with the channeling motion along 
a path which starts from the crystalline entrance and 
contains a given number of the undulator periods. The number of periods varies from 
zero (a de-channeling develops before the electron passes the first period) up to the 
maximum number of five periods (a primary channeling segment of the trajectory covers the 
entire CU length). 
For comparison, we have also calculated separate spectra produced by the non-accepted electrons 
and by the electrons that did not develop a channeling motion in the CU. We refer to 
Fig.~\ref{e-5micron-trajectories.fig} for the examples of trajectories from the 
different groups. 
The averaged radiation spectra for each group of the trajectories 
have a meaning of the spectra which would have been produced if all the simulated trajectories 
were of the kinds as these in the groups. They are shown in the top plot 
of Fig.~\ref{e-spectra-partial}. The bottom plot 
shows the absolute contributions of the spectra formed by the different groups of trajectories 
into the total simulated radiation spectrum.

The spectra in Fig.~\ref{e-spectra-partial} clearly display the lines 
resulting from oscillations in the projectile's motion. 
The line positions, being slightly different for different partial spectra, 
can be deduced from the total simulated spectrum as $1.1$~MeV and $8.7$~MeV 
for the undulator and channeling lines, respectively. 
The line intensities and widths also vary from one partial spectrum to another, however 
the more bending periods the electrons follow in the channeling mode upon entering the CU, 
the larger is an excess of the undulator line intensities over the channeling ones. 
The most intensive and narrow undulator lines are produced in the spectra 
normalized ``per a single trajectory'' which correspond to four and five full 
periods of the channeling motion (see the top plot in Fig.~\ref{e-spectra-partial}). These 
spectra, as well as the spectra for three and two full periods of the channeling motion, 
also display the lines of the second harmonics of the undulator radiation. 
The channeling lines are significantly broader than the undulator lines 
(notice the logarithmic scale for the radiation energies), implying that the channeling 
oscillations develop in the motion in a less harmonic manner than the undulator oscillations. 

Though the undulator lines formed by the trajectories with two to five 
full CU periods of the channeling motion are intensive in the partial spectra 
normalized per a single electron (top plot), they have low to moderate weights in the 
total simulated spectrum (bottom plot) as a result of small statistical wights of the 
corresponding channeling scenarios. The largest contribution to the total spectrum comes 
from the non-accepted trajectories. It is however remarkable, that the partial spectra 
produced by the non-accepted electrons display a prominent undulator line 
(see the red curves in the plots). The line clearly results from the re-channeling 
of the electrons and dominantly contributes the undulator line in total spectrum. 
Thus, the re-channeling, being enhanced by the periodical bending, plays an important role 
in producing a spectral undulator peak in the radiation from CU.
 
The effect of the radiation aperture on the spectral shape is studied in 
Fig.~\ref{e-5micron-vs-straight_dE.fig} where the spectra produced from CU 
are also compared with the spectra for the straight crystal. 
For both aperture values selected, $0.24$~mrad (top plot) and $4$~mrad (bottom plot), 
the maxima of the undulator and channeling spectral peaks 
are well separated from each other and display close intensities. Thus, the lines 
can be well resolved in an experiment. The peak intensities gain a factor of five with 
changing the aperture value from the smaller to the larger one. 
The channeling peak is suppressed in the radiation from CU as compared to the 
radiation from the straight crystal.   

Though the subject of our studies, in relation with the experiments at MAMI, 
is the channeling of electrons, it is instructive to consider also the channeling 
of positrons under the same conditions, in particular in the same CU. 
In Fig.~\ref{p-5micron-vs-straight_dE.fig}, we present the spectra computed from 
the corresponding simulations for the periodically bent and (for the reference) straight 
diamond samples. As the channeling motion of the positrons is more harmonic than that of 
the electrons, the spectra produced by the positrons display more intensive and narrow 
lines than the spectra for the electrons. In addition to the fundamental undulator peak 
in the radiation spectra, we encounter a sequence of the higher spectral harmonics. 
In the simulated spectra one can also distinguish the second spectral harmonics produced 
due to the channeling oscillations of the positrons, in addition to the well displayed 
fundamental channeling lines. For the smaller aperture value (top plot in the figure), 
the spectral lines have symmetric profiles typical for the radiation in the forward 
($\theta=0$) direction. For the larger aperture (bottom plot), the lines become broader 
towards softer radiation energies, as a result of dependence of the line positions 
on the radiation angle.  

\section{Conclusion}

We have presented the results of theoretical simulations on the channeling 
and radiation of the $855$~MeV electrons in the straight, arc-bent and periodically 
bent diamond (110) structures. A particular focus of the studies is on a 
large-amplitude long-period diamond CU used for the current experiments at MAMI. 
We have found that an enhanced re-channeling process develops in CU making the 
channeling very contrast to that in the arc-bent crystals. This process significantly 
contributes to the formation of the distinct spectral undulator lines in the 
radiation from CU. The spectral simulations for the positron beam of the same energy 
reveal the undulator lines even more pronounced as for the electron beam. 
We expect the results of our studies to be important for the current and future 
experiments with large-amplitude long-period CUs.

\begin{acknowledgement}
Financial support by the European Commission through the PEARL Project
within the H2020-MSCA-RISE-2015 call, GA 690991, is gratefully acknowledged.
\end{acknowledgement}

\end{document}